\documentclass{article}
\title{Fast, memory efficient low-rank approximation of SimRank}
\author{I.V. Oseledets \thanks{...}
        \and
        G.V. Ovchinnikov \thanks{The work is supported by RFBR grant 14-07-31297 mol\_a}
        }
\usepackage{graphicx}

\newenvironment{keywords}{{\bf Keywords:} }{}
\newcommand{\diag}{\mathop{\mathrm{diag}}}
\usepackage{amsmath}

\usepackage{color}
\usepackage{caption}
\usepackage{subcaption}
\usepackage{pgfplotstable}
\usepackage{booktabs}
\usepackage[ruled,vlined]{algorithm2e}

\begin{document}
\newpage
\maketitle
\begin{abstract}
    SimRank is a well-known similarity measure between graph vertices.  
    In this paper novel low-rank approximation of SimRank is proposed. 
    
\end{abstract}
\section{Introduction}
SimRank \cite{SimRank} is a topologically induced similarity measure between graph vertices which could be very useful in many applications, such as relation mining or document-by-document querying but, high computational $\mathcal{O}(n^3)$ and storage $\mathcal{O}(n^2)$ costs (here $n$ is number of vertexes in the graph) hinder it wider adoption. 
There have been numerous attempts at solving those issues.  
\cite{Li2010} proposes solution of Sylvester equation as an approximation of SimRank. 
\cite{He2012} proposes GPU-based solvers, and \cite{Lizorkin2010} describes some optimizations via threshold-sieving heuristics. 
While some of those techniques improve computational costs or propose hardware efficient methods none of the papers takes on improving memory consumption asymptotic. 
There is an empirical data from \cite{Cai2009} which suggests an existence of low-dimensional 'core' of SimRank for many real-world data, yet authors do not explore this idea. 
In this paper we propose to use low-rank matrices for efficient storage and computation of SimRank at the price of slightly lower accuracy of SimRank scores.

\section{Definitions}
In the foundation of SimRank definition lies idea that "two objects are similar if they are referenced by the similar objects". 
Next, it is proposed that object similarity lies between $0$ and $1$ with object being maximally similar to itself with similarity $1$.
Similarity between objects $a$ and $b$, $s(a,b)$ is defined in the following way:
$$
s(a,b) = \left\{\begin{array}{l}
1,\; a=b,\\
0,\; $if$\; I(a) = \emptyset \; $or$\; I(b) = \emptyset,\\ 
\displaystyle \frac{c}{|I(a)| |I(b)|}\sum_{i=1}^{|I(a)|}\sum_{j=1}^{|I(b)|}s(I_i(a),I_j(b)),\; $otherwise$,
      \end{array}
\right.
$$
here $I(v)$ is the set of in-neighbours of vertex $v$, constant $c \in (0,1)$. Writing $n^2$ equations (one equation for each pair of vertices) gives us a system 
with unique solution \cite{SimRank}.  

One can find a solution of such system by using the following iterative process:

\begin{equation}
\label{si1a}
R_0(a,b) = \left\{\begin{array}{l}
1,\; a=b,\\
0,\; $otherwise$,
      \end{array}
\right.
\end{equation}
\begin{equation}
\label{si1b}
R_{k+1}(a,b) = \left\{\begin{array}{l} 
    1,\; a=b,\\
    0,\; $if$\; I(a) = \emptyset \; $or$\; I(b) = \emptyset,\\ 
    \displaystyle \frac{c}{|I(a)||I(b)|}\sum_{v \in I(a)}\sum_{w \in I(b)}R_k(w,v),\; $otherwise$.
    \end{array}
    \right.
\end{equation}
It is shown in \cite{SimRank} that $R_k(a,b)$ converges to $s(a,b)$.

The iterative process (\ref{si1b}) can be written in the matrix form as follows:
\begin{equation}
\label{si2}
S_{k+1} = c W^TS_{k}W -c\diag(W^TS_{k}W) + I,\, S_0 = I,
\end{equation}
here $W$ is the adjacency matrix $A$ normalized by columns $W = AD^{-1}$
with $D$ being a diagonal matrix with 
$$
D_{i,i} = \sum_{j=1}^{n}A_{i,j}.
$$
\section{Low-rank approximation}
Due to SimRank itself being a symmetric matrix (which can be seen from (\ref{si2})) we choose symmetric form for the low-rank approximation
\begin{equation}
\label{low_rank_form}
\tilde{S} = I + UDU^T,
\end{equation}
where $I$ is an identity matrix of $n$-th order, $U$ and $D$ are correspondingly orthonormal and diagonal matrices of sizes $n \times m$ and $m \times m$.

Instead of the iterative process (\ref{si2}) converging to the precise SimRank matrix $S$ we consider the following iterative process:
\begin{equation}
\label{low_rank_iter}
\tilde{S}_{k+1} = W^T\tilde{S}_kW - \diag(W^T\tilde{S}_kW) + I.
\end{equation}
Substituting (\ref{low_rank_form})  into (\ref{low_rank_iter}) with arbitrary initial matrices $U_0$ and $D_0$ gives us the following equation for $U_{k+1}$ and $D_{k+1}$:
$$
I + U_{k+1}D_{k+1}U^T_{k+1} = W^T(I + U_kD_kU^T_k)W - \diag(W^TW + W^TU_kDU^T_kW ) + I,
$$
from which we derive 
$$
U_{k+1}D_{k+1}U_{k+1}^T = M_{k},
$$
where $M_{k} = W^T(I + U_kU_k^T)W -\diag(W^TW + W^TU_kU^T_kW )$.
Calculation of matrices $U_{k+1}$ and $D_{k+1}$ is based on the eigenvalue decomposition of matrix $M$: 
$$
M_{k} = \tilde{U}\tilde{D}\tilde{U}^T, 
$$
here $\tilde{U}, \tilde{D}$ are matrices of eigenvectors and eigenvalues correspondingly, and $\tilde{D}$ is ordered from greatest to least along its diagonal. Using a priori chosen approximation rank $r$ we from matrices $U_{k+1}$ and $D_{k+1}$:
$$
U_{k+1} = \tilde{U}_r, D_{k+1} = \tilde{D}_r,
$$
here $\tilde{U}_r$ is matrix of the first $r$ eigenvectors and $\tilde{D}_r$ is a matrix of first $r$ eigenvalues. 
Convergence analysis is complicated due to $k$-rank projection operation not being Lipschitz. 

\section{Advantages over existing methods}
The proposed method has memory requirements $O(n\max(r,d))$ and computational complexity $O(rkn^2)$. The iterative algorithm from the original paper \cite{SimRank} has memory requirements $O(n^2)$ and computational complexity $O(kn^2d^2)$ which is $O(kn^4)$ in worst-case scenario. A number of attempts was made to speed-up SimRank computation. \cite{Lizorkin2010} proposes three optimization strategies: skipping computation of similarity between some node pairs, optimizations of SimRank iterative process and pruning. In the end it improves worst-case computational complexity of algorithm to $\min(O(nl), O(n^3 / \log_2(n))$, with $l$ denoting the number of object-to-object relationships. The memory requirements are the same as with the original paper, namely 
$O(n^2)$.

\begin{figure}[h]
\includegraphics[width=0.8\textwidth]{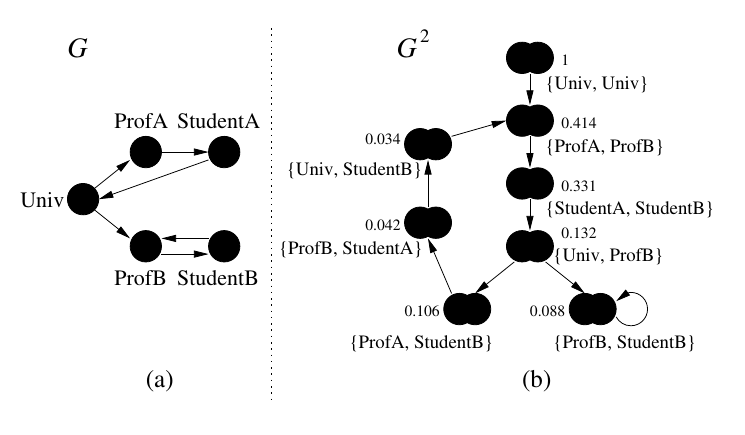}
\caption{Example graph from original SimRank paper. a) Graph b) Corresponding SimRank scores}
\label{fig:graph}
\end{figure}

Another approach, based on the approximation of SimRank by Sylvester equation is pioneered in \cite{Li2010}. While this approach provides nice and well understood algebraic equation with already available solvers implemented in all popular programming languages, it requires computation of $W^{-1}$ which is a full matrix, while $W$ is usually sparse. There are number of works extending on this idea, for example \cite{Onizuka2013} proposes low-rank approximation to Sylvester equation or \cite{Yu2014} proposes low-rank approximation for SimRank equation written in the form of discrete Lyapunov equation. It should be noted that those equations are in fact only approximation to SimRank. It can be easily seen from an example graph provided in original paper \cite{SimRank}. The adjacency matrix  is 
$$
W = \left(\begin{array}{ccccc} 0 & 1.0 & 0 & 1.0 & 0\\ 0 & 0 & 1.0 & 0 & 0\\ 1.0 & 0 & 0 & 0 & 0\\ 0 & 0 & 0 & 0 & 1.0\\ 0 & 0 & 0 & 1.0 & 0 \end{array}\right).
$$
The result for original iterative simrank algorithm gives the following SimRank matrix: 
$$
\left(\begin{array}{ccccc} 1.0 & 0 & 0 & 0.1323 & 0.03388\\ 0 & 1.0 & 0 & 0.4136 & 0.1059\\ 0 & 0 & 1.0 & 0.04235 & 0.3308\\ 0.1323 & 0.4136 & 0.04235 & 1.0 & 0.08822\\ 0.03388 & 0.1059 & 0.3308 & 0.08822 & 1.0 \end{array}\right)
$$
The result of the Lyapunov equation based approximation from \cite{Li2010} gives:
$$
\left(\begin{array}{ccccc} 1.0 & 3.013\cdot 10^{-15} & 8.797\cdot 10^{-16} & 0.1323 & 0.03388\\ 6.25\cdot 10^{-16} & 1.0 & 2.796\cdot 10^{-15} & 0.4136 & 0.1059\\ 3.2\cdot 10^{-15} & 6.729\cdot 10^{-16} & 1.0 & 0.04235 & 0.3308\\ 0.1323 & 0.4136 & 0.04235 & 0.5399 & 0.08822\\ 0.03388 & 0.1059 & 0.3308 & 0.08822 & 0.632 \end{array}\right)
$$
As we can see even for such small and simple example the discrepancy not negligible: the $S_{4,4}$ and $S_{5,5}$ significantly differ from $1$. On slightly more complex tests non-diagonal elements will differ, too.
For example, we can modify this graph adding an edge between StudentA and StudentB (see the Figure~\ref{fig:graph}) which gives us the following adjacency matrix:
$$
W = \left(\begin{array}{ccccc} 0 & 1.0 & 0 & 1.0 & 0\\ 0 & 0 & 1.0 & 0 & 0\\ 1.0 & 0 & 0 & 0 & 1.0\\ 0 & 0 & 0 & 0 & 1.0\\ 0 & 0 & 1.0 & 1.0 & 0 \end{array}\right)
$$

The result for original iterative SimRank algorithm: 
$$
\left(\begin{array}{ccccc} 1.0 & 0.1809 & 0.2262 & 0.1993 & 0.5523\\ 0.1809 & 1.0 & 0.2933 & 0.6209 & 0.1702\\ 0.2262 & 0.2933 & 1.0 & 0.3807 & 0.2721\\ 0.1993 & 0.6209 & 0.3807 & 1.0 & 0.1744\\ 0.5523 & 0.1702 & 0.2721 & 0.1744 & 1.0 \end{array}\right),
$$
and result of algorithm from \cite{Li2010}:
$$
\left(\begin{array}{ccccc} 0.5653 & 0.08779 & 0.1097 & 0.09898 & 0.2538\\ 0.08779 & 0.6522 & 0.1366 & 0.3276 & 0.08348\\ 0.1097 & 0.1366 & 0.4566 & 0.1778 & 0.1377\\ 0.09898 & 0.3276 & 0.1778 & 0.5073 & 0.08661\\ 0.2538 & 0.08348 & 0.1377 & 0.08661 & 0.4639 \end{array}\right).
$$
As one can see the differences are too large to be ignored.

\section{Speeding up with probabilistic spectral decomposition} 

Spectral decomposition of the matrix $M$ could be replaced by its probabilistic approximation from work \cite{HMT}. Here is a short description of used algorithm, provided for readers convenience. 

\begin{algorithm}[H]
 \SetAlgoLined
 \KwData{Symmetric matrix $M$, approximation rank $r$, oversampling parameter $p$}
 \KwResult{Approximate eigenvalue decomposition $M \approx U\Lambda U^T$}
    1. Select approximation rank $r$ and oversampling parameter $p$\; 
    2. Generate $n \times (r + p)$ matrix $Z$ with elements drawn from standard Gaussian distribution\;
    3. Form $n \times (r+p)$ matrix $Y = MZ$\;
    4. Construct an $n \times (r+p)$ matrix $Q$ whose columns form an orthonormal basis for the range of $Y$\;
    5. Form $ (r+p) \times (r+p)$ matrix $B = Q^TMQ$\;
    6. Compute spectral decomposition $B = V^T\Lambda V$\;
    7. Use $QV$ as approximation of eigenvectors of matrix $M$\;
 \caption{Probabilistic spectral decomposition}
\end{algorithm}
For more detailed information about probabilistic approaches for low-rank approximation we direct reader to \cite{HMT}.

Using this our final algorithm can be formulated as follows:

\begin{algorithm}[H]
 \SetAlgoLined
 \KwData{Column normalized adjacency matrix $W$, approximation rank $r$, oversampling parameter $p$, number of iterations $k$}
 \KwResult{Matrices $U$ and $D$ such $S \approx I + UDU^T$}
 set initial values for $U$, $D$ \;
 \For{$i$=$1\dots k$}{
 $A_1 \leftarrow W^TWZ$\;
 $A_2 \leftarrow W^TUDU^TWZ$\;
 $A_3 \leftarrow diag(A_1+A_2)Z$\;
 $U,D \leftarrow$ probabilistic spectral decomposition of $A_1+A_2-A_3$\;
 } 
 \caption{Low-rank SimRank approximation}
\end{algorithm}

We can also have a better approximation of SimRank in the form 
\begin{equation}
\label{one_more}
S \approx I + W^TW + W^TUDU^TW - W^T diag (W^TW + W^TUDU^TW) W,
\end{equation}
which is essentially one iterative step done as in original iterative algorithm (\ref{si2}) with approximative SimRank used as initial guess. While this form will result in dense matrix this is form can be effectively used for querying.
\section{Numerical experiments}

The proposed method provides fixed memory and computation complexity, but the question about the approximation quality is still open. While no amount of numerical experiments can replace good theoretical estimates it nevertheless sheds some light on proposed method behaviour. All experiments are done in MATLAB 2014a.

\begin{figure}[t]
\centering
\begin{subfigure}{0.49\linewidth} \centering
        \includegraphics[width=1.0\textwidth]{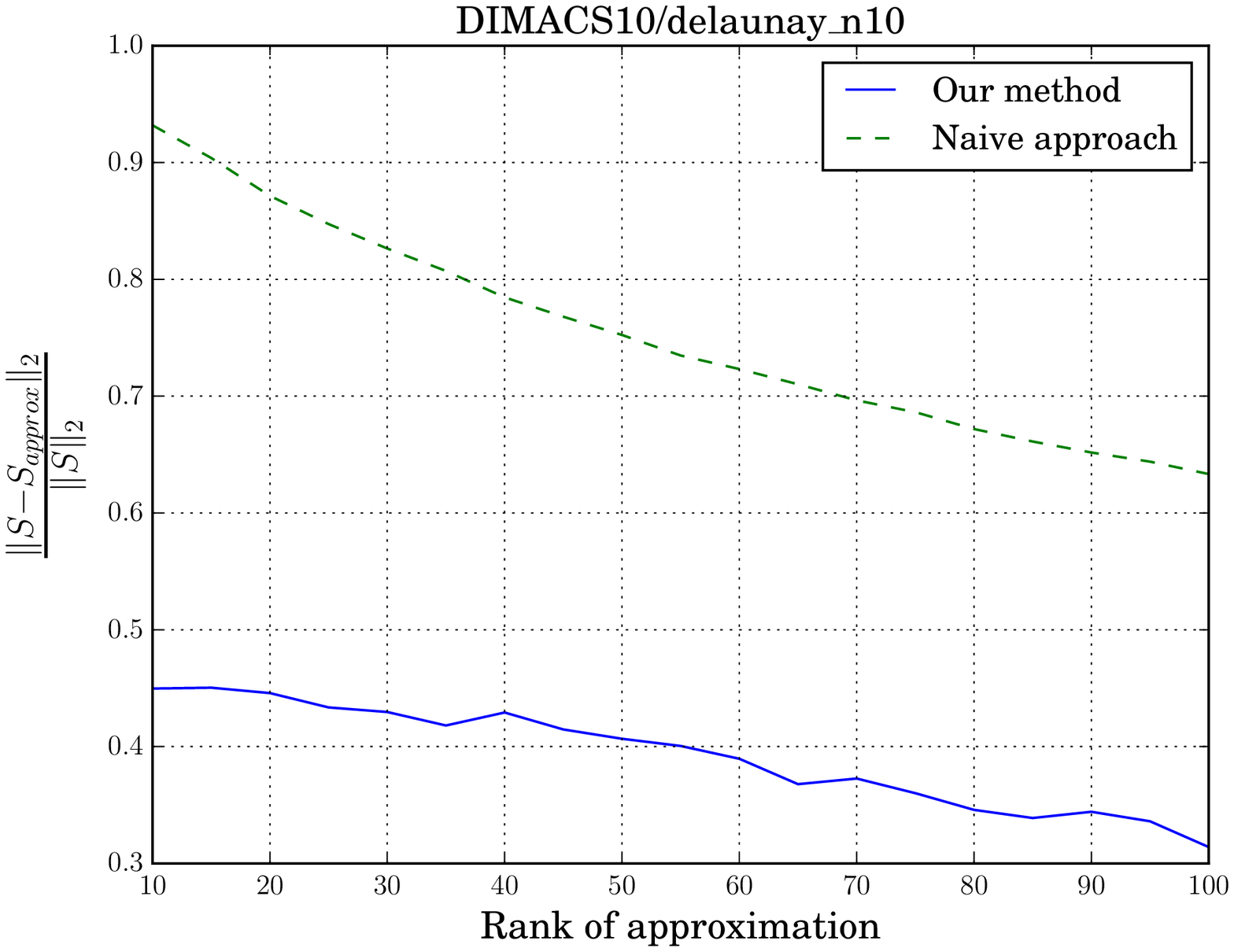}
\end{subfigure}
\begin{subfigure}{0.49\linewidth}
  \centering
  \includegraphics[width=1.0\textwidth]{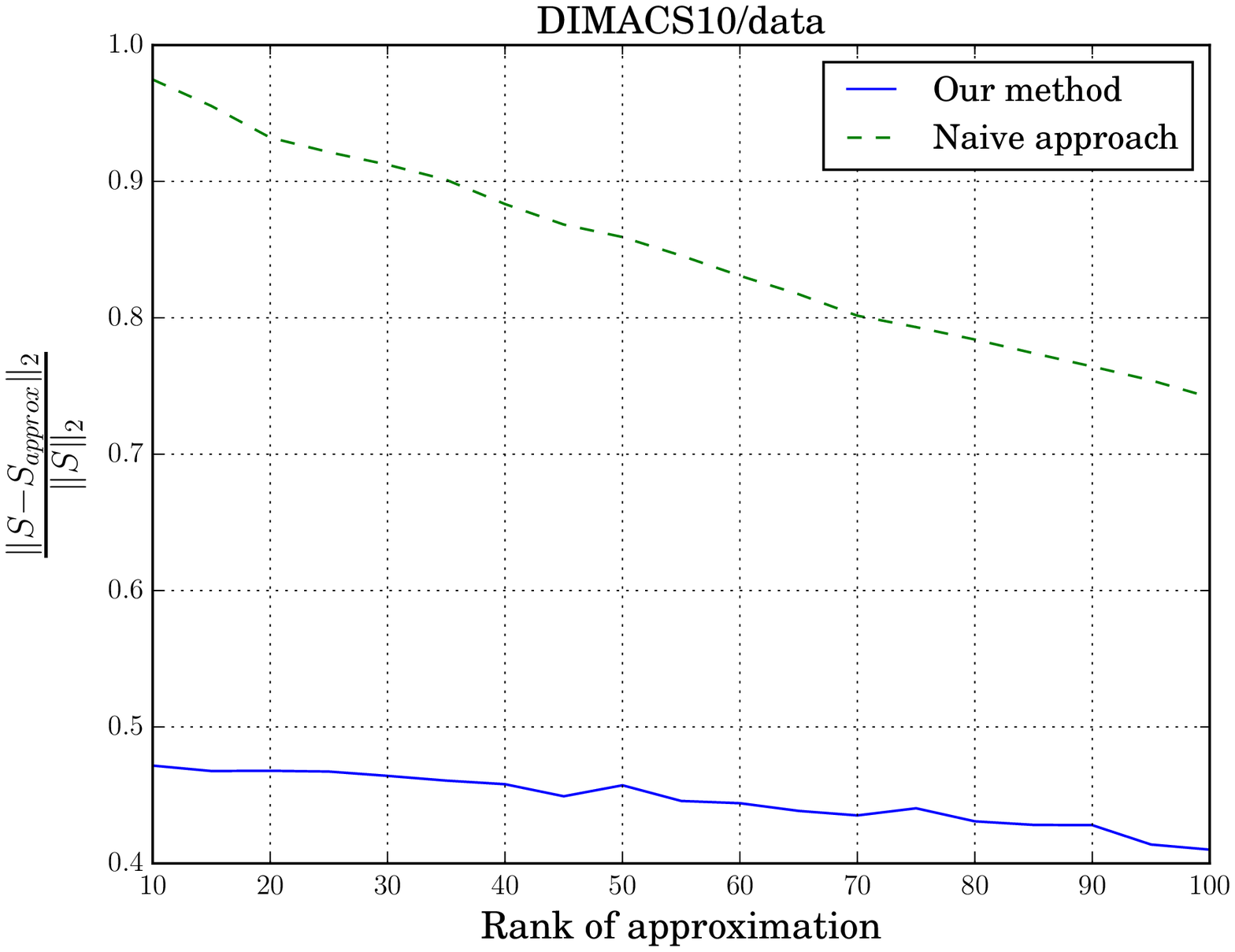}
\end{subfigure}
\caption{Dependence of relative error on approximation rank}
\label{fig:rel}
\end{figure}

\begin{figure}[t]
\centering
\begin{subfigure}{0.49\linewidth} \centering
        \includegraphics[width=1.0\textwidth]{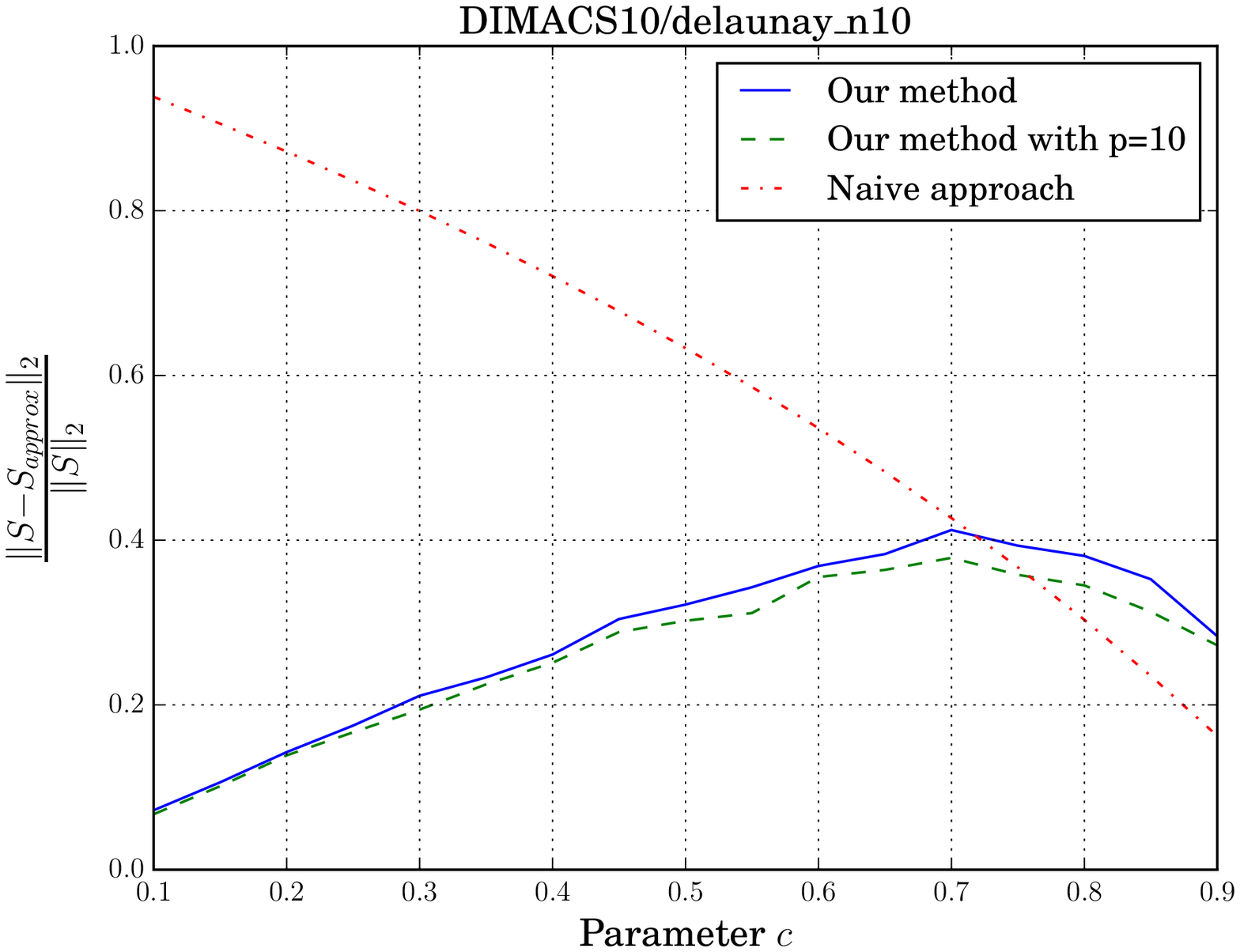}
\end{subfigure}
\begin{subfigure}{0.49\linewidth}
  \centering
  \includegraphics[width=1.0\textwidth]{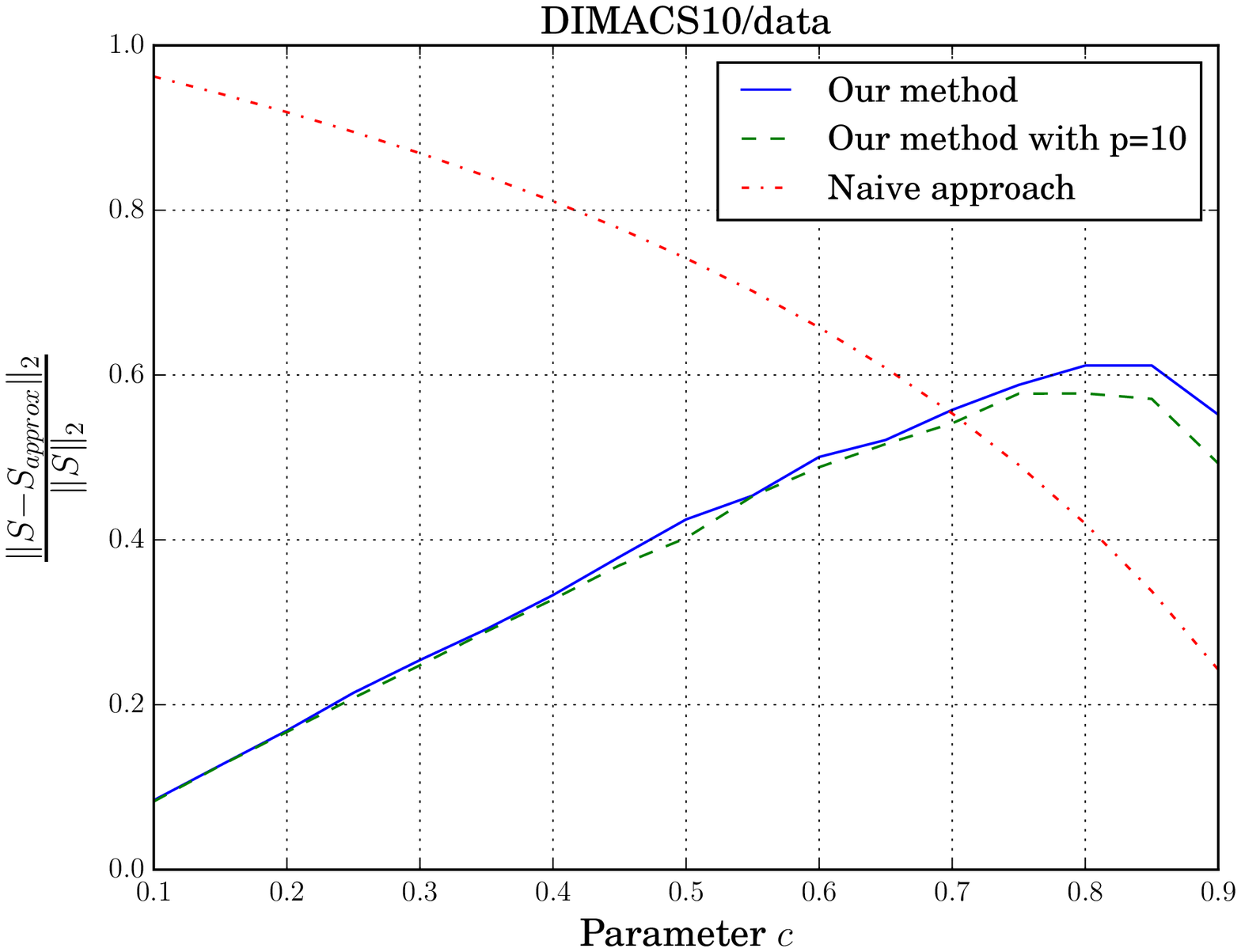}
\end{subfigure}
\caption{Dependence of relative error on $c$}
\label{fig:c}
\end{figure}

In this section we present our experimental study of proposed algorithm on a graphs from DIMACS10 Challenge Collection \footnote{https://www.cise.ufl.edu/research/sparse/matrices/DIMACS10/}. First we studied dependence of relative error of our method on approximation rank. 
For comparison we used best rank $r$ approximation of the SimRank matrix. We call this method naive because it does not take into account structure of the SimRank matrix. Methods based on Sylvester or Lyapunov equation usually 
employ some kind of low-rank approximation and, if those equations would describe SimRank precisely, it would be their upper limit. 
As can be seen from Figure~\ref{fig:rel} the proposed method provides superior accuracy. For those experiments $c=0.5$ was selected and no oversampling ($p=0$) was used. 
The second part of the experiments shows dependence of relative error on $c$ with and without oversampling.  
Based on estimations from \cite{HMT} $p=10$ is chosen. As we can see from Figure~\ref{fig:c}    
our method provides better approximation for most part of values of $c$.  As expected oversampling provides some improvement.

\begin{tabular}{|rrrrrrr|}\hline 
        name & $n$ & rank & nnz   & $nnz/n$  &corr & err \\ \hline 
        chesapeake & 39 & 3 & 340 &  8.72  &0.57 &     0.12 \\ 
        data & 2851 & 285 & 30186 &  10.59 &0.82 &     0.42 \\ 
 delaunay n10 & 1024 & 102 & 6112 &  5.97  &0.67 &     0.30 \\ 
 delaunay n11 & 2048 & 204 & 12254 & 5.98  &0.67 &     0.41 \\ 
 delaunay n12 & 4096 & 409 & 24528 & 5.99  &0.68 &     0.60 \\ 
 delaunay n13 & 8192 & 819 & 49094 & 5.99  &0.68 &     0.82 \\ 
           uk & 4824 & 482 & 13674 & 2.83  &0.24 &     0.66 \\ 
        vsp data and seymourl & 9167 & 916 & 111732 &  12.19 & 0.89 &     0.39 \\ \hline 
\end{tabular}

Here $nnz$ is the number of nonzero elements in the adjacency matrix, $nnz/n$ is the average degree of vertex, corr is correlation between $S$ and $\tilde{S}$ with diagonals removed and
$$
\mbox{err} = \| \frac{S}{\|S\|_2} - \frac{\tilde{S}}{\|\tilde{S}\|_2} \|_2.
$$
From this table is clearly seen that the accuracy of our method depends on the connectivity of the given graph, the higher the connectivity the greater accuracy will be.

\subsection{Experiment with Wikipedia}
We used Simple English Wikipedia corpus to find semantic relatedness between concepts. We had $150495$ entities(order of our matrix) which was a slightly higher than the number of articles in the given wiki at the time of writing, because some of those entities are redirection stubs. We used undirected graph representation of inter-wiki links which gave us $4454023$ non zero elements in adjacency matrix. For graph that large direct computation of SimRank is infeasible. We used the following parameters for the experiment: $c = 0.3$, $\mbox{rank} = 6000$, no-oversampling and did ten iterations. 
While original paper \cite{SimRank} suggests $c=0.8$ later it was suggested to use $c=0.6$ for better results \cite{Lizorkin2010} and we choose $c=0.3$ because it gave us subjectively better results.  In experiment we used virtual server (VZ container) with 16 CPU cores and 100GiB RAM available (host node has 32 cores: 4 CPUs, each is 8-core AMD Opteron Processor 6272, 128GiB RAM). With this setup computations took roughly 40 hours.

Some examples provided in the table below. The first row is the word for which most similar words were queried, then in each of the columns most similar words are listed ordered by their SimRank score. The scores themselves would take too much space (they differ in 4-th or 5-th significant figures) and hence are omitted. 

\vspace{5mm}
\begin{figure}[h]
\includegraphics[width=1.0\textwidth]{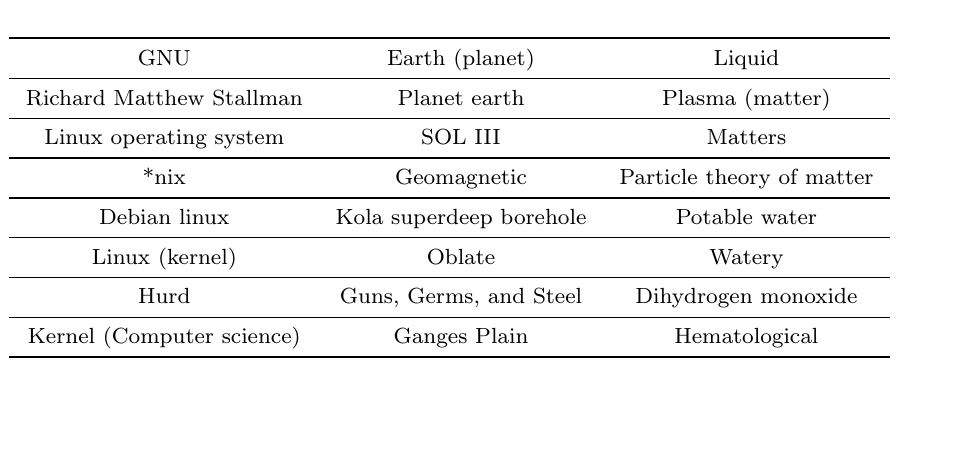}
\end{figure}

Those results look very promising, despite small approximation rank (compared to matrix order). Two factors contribute to this: high average vertex degree of $\approx 29.6$ and empirically observed localization of errors in more dissimilar items. Investigation of both factors requires further work.

\section{Acknowledgements}
We are particularly grateful for the valuable technical assistance given by A.A. Sadovsky and D.A. Podlesnykh.

\bibliographystyle{plain}
\bibliography{references}

\end{document}